\documentclass[showpacs,amsmath,amssymb,aps,showkeys,floatfix,prd,a4paper]{revtex4}
\usepackage{graphicx,epstopdf}
\usepackage{xcolor}
\usepackage{subfigure}
\usepackage{dcolumn}
\usepackage{bm}
\usepackage{epsfig}
\usepackage{amsfonts}
\usepackage{amssymb,amscd}
\def\lsim{\raise0.3ex\hbox{$<$\kern-0.75em\raise-1.1ex\hbox{$\sim$}}}
\def\gsim{\raise0.3ex\hbox{$>$\kern-0.75em\raise-1.1ex\hbox{$\sim$}}}

\newcommand{\be}{\begin{equation}}
\newcommand{\ee}{\end{equation}}

\def\beq{\begin{equation}}
\def\eeq{\end{equation}}
\def\beqa{\begin{eqnarray}}
\def\eeqa{\end{eqnarray}}

\newcommand{\ba}{\begin{eqnarray}}   
\newcommand{\eea}{\end{eqnarray}}

\def\gappeq{\mathrel{\rlap {\raise.5ex\hbox{$>$}}
{\lower.5ex\hbox{$\sim$}}}}
\def\lappeq{\mathrel{\rlap{\raise.5ex\hbox{$<$}}
{\lower.5ex\hbox{$\sim$}}}}
\def\Toprel#1\over#2{\mathrel{\mathop{#2}\limits^{#1}}}

\begin{document}
\begin{flushright}
\vskip1cm
\end{flushright}

\title{ Impact of the inelastic proton -- nucleus cross section on the prompt neutrino flux}

\author{A. V. Giannini}
\affiliation{ Instituto de F\'{\i}sica, Universidade de S\~{a}o Paulo\\
C.P. 66318,  05315-970 S\~{a}o Paulo, SP, Brazil}

\author{V. P. Gon\c{c}alves}
\affiliation{High and Medium Energy Group,\\ Instituto de F\'{\i}sica e Matem\'atica,\\  Universidade Federal de Pelotas (UFPel)\\
Caixa Postal 354,  96010-900, Pelotas, RS, Brazil.}

\begin{abstract}
The description of the inelastic proton -- nucleus cross section at very high energies is still an open question. The current theoretical uncertainty  has direct impact on the predictions of the cosmic ray and neutrino physics observables. In this paper we consider different models for the treatment of $\sigma_{inel}^{pA}$, compare its predictions at ultrahigh cosmic ray energies and estimate the prompt neutrino flux at the neutrino energies that have been probed by the IceCube Observatory. We demonstrate that depending of the model used to describe   $\sigma_{inel}^{pA}$, the predictions for the prompt neutrino flux can differ by a factor of order of three. Such result demonstrate the importance of a precise measurement of the inelastic proton -- nucleus cross section at high energies.
 \end{abstract}



\date{\today}

\maketitle

\begin{figure}[t]
\centering
\includegraphics[scale=0.75]{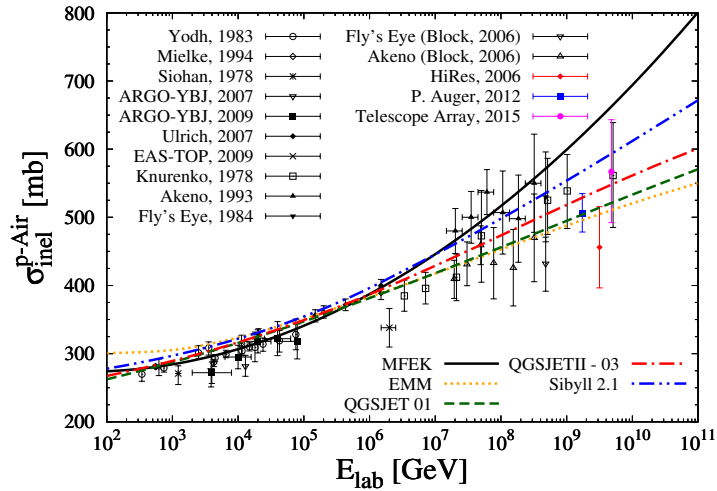}
\caption{Energy dependence of the inelastic proton - Air cross section predicted by the distinct phenomenological models discussed in the text. }
\label{fig:1}
\end{figure}

The description of the conventional and prompt atmospheric neutrino fluxes, produced by  cosmic-ray interactions with nuclei in the 
atmosphere  of the Earth,  has been the theme of a series of studies in the last years \cite{calc_recentes,rojo1,anna2,vicantoni}. Such analysis were strongly motivated by the detection of astrophysical neutrinos  by the IceCube Observatory 
\cite{IceCube_Science,Aartsen:2014gkd,Aartsen:2016xlq}. In order to determine the cosmic neutrino flux, it is fundamental to have a precise knowledge of the atmospheric neutrino flux (For a recent review see e.g. Ref. \cite{ahlers}). Currently, the description of  the prompt contribution is a subject of intense debate, since it strongly depends on the modelling of the heavy quark production at large energies and very forward rapidities, beyond those probed at the LHC  \cite{vicantoni}. Although the  LHC data on the prompt heavy quark  cross sections       
(see e.g. Refs.~\cite{Aaij:2013mga,Aaij:2015bpa}) helped us to improve  
the description of  heavy meson production at forward rapidities and   
significantly reduced some of the theoretical uncertainties, 
the predictions obtained by different groups can still differ by a 
factor $\ge 2$ depending on the treatment of heavy quark production at 
high energies and of the QCD dynamics at small values of 
the Bjorken - $x$ variable \cite{calc_recentes,rojo1,anna2,vicantoni}. These previous studies have mainly focused  in the calculation of the prompt neutrino flux considering different approaches for the factorization of the heavy quark cross section, distinct parametrizations for the parton distribution functions as well different models for the primary incident nucleon flux. Another important ingredient in the calculation of the prompt neutrino flux at high energies is the inelastic proton - Air cross section ($\sigma_{inel}^{pAir}$), which determines the magnitude of the $Z$ - moments and, consequently, the evolution of the hadronic cascades in the atmosphere.  The modelling of $\sigma_{inel}^{pA}$  has been discussed by several authors during the last years (See e.g. Refs. 
\cite{Ulrich,Portugal:2009xc,dEnterria:2011twh,Fagundes:2011hv}), with its predictions  at high energies {being} largely distinct (See Fig. \ref{fig:1}). 
In this paper we { will} estimate the impact of these different models for
$\sigma_{inel}^{pAir}$ on the prompt neutrino flux at high neutrino energies, as those probed by the IceCube Observatory and future neutrino telescopes. As we will demonstrate in what follows,  the current uncertainty associated to the treatment of the proton - Air cross section at high energies is a factor of order of 3, independent of the model used to describe the primary nucleon flux. Such result demonstrate the importance of a precise measurement of $\sigma_{inel}^{pAir}$ at Cosmic Ray energies.

Initially let's present a brief review of the formalism used to estimate the prompt neutrino flux and refer the reader to Refs. \cite{anna2,vicantoni} for more details. As in Ref. \cite{vicantoni}, we will calculate the prompt neutrino flux  
using the semi-analytical $Z$-moment approach, proposed many years ago in Ref.~\cite{gaisser_book} and
discussed in detail e.g.~in Refs.~\cite{ingelman,ers,rojo1}.
In this approach, a set of coupled cascade 
equations for the nucleons, heavy mesons and leptons (and their antiparticles) 
fluxes is solved, with the equations being expressed in terms of the nucleon-to-hadron
($Z_{NH}$), nucleon-to-nucleon ($Z_{NN}$), hadron-to-hadron  ($Z_{HH}$) 
and hadron-to-neutrino ($Z_{H\nu}$) $Z$-moments. 
These moments are inputs in the calculation of the prompt neutrino flux
associated with the production of a heavy hadron $H$ and its decay into a
neutrino $\nu$ in the low- and high-energy regimes. 
As a example we present the definition of the heavy hadron $Z$-moment, which can be expressed as follows 
\begin{eqnarray}
Z_{pH} (E) =  \int_0^1 \frac{dx_F}{x_F} \frac{\phi_N(E/x_F)}{\phi_N(E)} 
\frac{1}{\sigma_{inel}^{pAir}(E)} \frac{d\sigma_{pAir \rightarrow H}(E/x_F)}{dx_F} \,\,,
\label{eq:zpH}
\end{eqnarray}
where $E$ is the energy of the produced particle (heavy hadron), $x_F$ 
is the Feynman variable, $\phi_N$ is the primary cosmic ray flux, $\sigma_{inel}^{pAir}$ is the inelastic proton-Air
cross section and 
$d\sigma/dx_F$ is the differential cross section for  heavy hadron production.  In what follows we will focus on 
vertical fluxes  and will 
assume that the cosmic ray flux $\phi_N$  can be described by a broken 
power-law (BPL) spectrum \cite{prs} or by the H3a spectrum proposed in Ref. \cite{gaisser}, with the incident flux being
represented by protons.  As in Ref. \cite{vicantoni},  we will assume that the charmed 
hadron $Z$-moments can be expressed in terms of the charm $Z$-moment 
as follows: $Z_{pH} = f_H \times Z_{pc}$, where $f_H$ is the fraction of charmed
particle which emerges as a hadron $H$, which will be assumed to be:  
$f_{D^0} = 0.565$, $f_{D^+} = 0.246$, 
$f_{D_s^+} = 0.080$ and $f_{\Lambda_c} = 0.094$ \cite{ers}.
Moreover, we will disregard nuclear effects assuming that $\sigma(pAir \rightarrow c\bar{c}) = 14.5 \times \sigma(pp \rightarrow c\bar{c})$. We will use the collinear factorization formalism to describe the heavy quark production and consider the CT14 parametrization \cite{ct14} to describe the quark and gluon distributions in the proton.
All other moments have been estimated as discussed in detail in Ref. \cite{vicantoni}.

  We are interested in the calculation of the prompt neutrino flux at high energies. One have that the production of neutrinos at a given neutrino 
energy, $E_{\nu}$, is determined by collisions of cosmic rays with nuclei in
the atmosphere at energies that are a factor of  100-1000 larger. Therefore,  the magnitude of the prompt neutrino flux measured  in the kinematical range that is probed by the IceCube Observatory and future neutrino telescopes is directly 
associated to the  the modelling of the inelastic cross section [See Eq. (\ref{eq:zpH})]. As briefly pointed out above, the treatment of this quantity at high energies is still an open question. The measurement of the inelastic proton-Air cross section performed by the Pierre Auger collaboration~\cite{Collaboration:2012wt} at $\sqrt{s} = 57$ TeV helped to improve our understanding of the hadronic interactions at high energies and constrain its description.  In what follows we will consider some phenomenological models commonly used in the literature and  the corresponding prompt neutrino flux will be estimated. In particular, we will consider 
 four different hadronic interaction models that derive the  inelastic proton-Air cross section using the Glauber formalism~\cite{Glauber:1970jm}. All of them are based on the eikonal representation but differ in the way the eikonal functions are constructed.
In the QGSJET01 model~\cite{Kalmykov:1997te} all eikonal functions are described via independent Pomeron exchanges, following the Gribov's reggeon framework~\cite{Gribov:1968fc,Gribov:1968jf}. An improved version of this model, denoted QGSJETII.03 model, was proposed in Ref. ~\cite{Ostapchenko:2005nj}, which differs from its predecessor by including non-linear effects, which are described in terms of Pomeron-Pomeron interactions in the employed framework. Such effects are important in the description of the hadronic interactions at high energies and small impact parameters. On the other hand, the Sibyll 2.1 model~\cite{Ahn:2009wx} is based on a two-channel (``soft" + ``hard") eikonal function: while the energy dependence of the ``soft" channel is modelled as a sum of two power laws, one related to Pomeron exchange and another related to Reggeon exchange, as in Regge theory~\cite{Donnachie:1992ny}, the energy dependence of the ``hard" component is modelled as a ``minijet" model \footnote{``Minijets" are jets with small transverse momentum, where perturbative calculations can be carried out, but smaller than a typically reconstructed jet at high energy colliders. Moreover, the minijet model neglects hard interactions with $p_{T} < p_{T}^{min}$ where the physics is assumed to be ``soft".}. In this model, the steep increasing  of the parton distributions at low values of Bjorken-$x$ ({\it i}. {\it e}. high energies) in the colliding hadrons leads to the increase of the hadronic cross section. In addition, we also have considered the QCD-inspired model from ref.~\cite{Giannini:2013jla} (denoted EMM hereafter) which also parametrizes the ``hard" channel as a minijet model (employing newer parton distribution functions) but handle the ``soft" channel and the $p_{T}^{min}$ different from Sibyll: while both quantities increase with the energy in the  Sibyll 2.1, Ref.~\cite{Giannini:2013jla} assumes a constant $p_{T}^{min}$ and a soft channel that is practically energy-independent, so that the hadronic cross section increases only due to hard (partonic) interactions. Finally, for completeness, we also consider the analytical parametrization for $\sigma_{inel}^{pAir}$  proposed in Ref. \cite{Mielke:1994un}, denoted by MFEK,  which has been used in early estimates of prompt neutrino fluxes. 

In Fig. \ref{fig:1} we present a comparison between the predictions of the phenomenological models with the experimental data for the inelastic proton - Air cross section. We have that its predictions are similar at low values of  the proton energy in the laboratory frame ($E_{lab}$), with the EMM one overestimating the data in this kinematical regime and the QGSJETII.03 being a lower bound. On the other hand, at large energies they differ by $\approx 40 \%$, with the MFEK (EMM) prediction being an upper (lower) bound.
 Although the experimental uncertainty is still large, one have that P. Auger (2012) data is not described by the MFEK and Sybill 2.1 models. It is important to emphasize that the more recent version of the Sybill Monte Carlo implies smaller values for $\sigma_{inel}^{pAir}$ at large energies, with its predictions being similar to those of the QGSJET model. 
In what follows we will estimate the impact of these differences on the predictions for the prompt neutrino flux.

\begin{figure}[t]
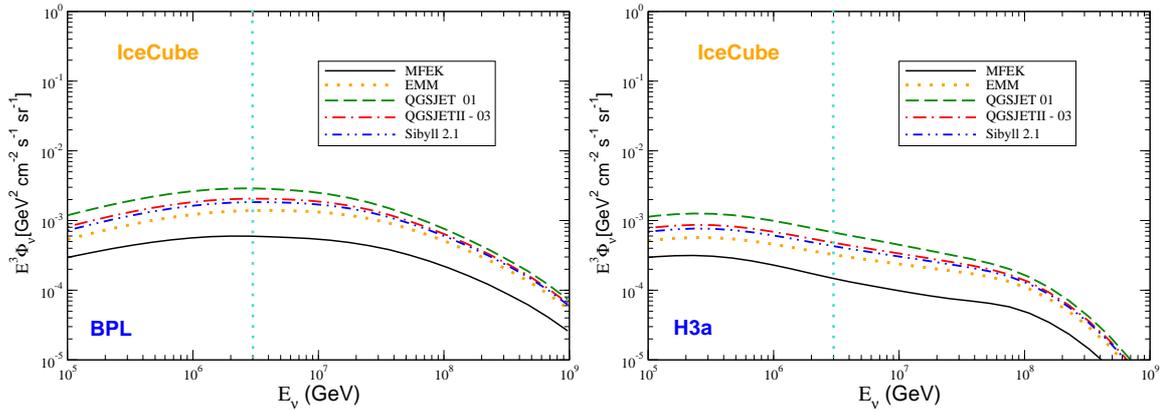

\begin{center}
\includegraphics[width=0.42\textwidth]{comp_sigpA_BPL.eps}
\includegraphics[width=0.42\textwidth]{comp_sigpA.eps}
\end{center}
\caption{Energy dependence of the prompt neutrino flux, normalized by a factor $E_{\nu}^3$, calculated assuming the BPL (left) and H3a (right) models for the primary cosmic ray flux and different models for the inelastic proton - Air cross section.
}
\label{fig:flux}
\end{figure}

\begin{figure}[t]
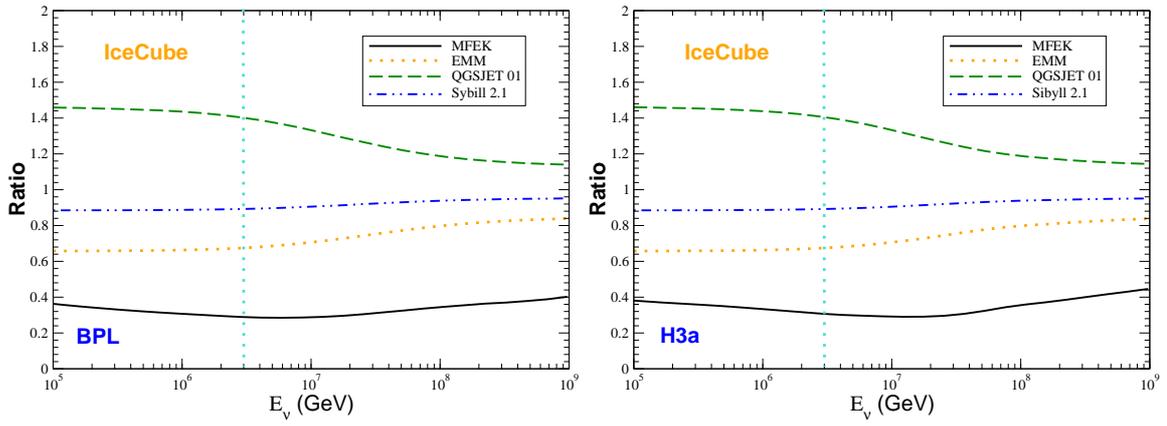

\begin{center}
\includegraphics[width=0.42\textwidth]{ratio_BPL.eps}
\includegraphics[width=0.42\textwidth]{ratio_H3a.eps}
\end{center}
\caption{Energy dependence for the ratio between the predictions of different models 
 for the inelastic proton - Air cross section and the QGSJETII-03 one, calculated assuming the BPL (left) and H3a (right) models for the primary cosmic ray flux.
}
\label{fig:ratio}
\end{figure}

In Fig. \ref{fig:flux} we present our predictions for the energy dependence of the prompt neutrino flux, normalized by a factor $E_{\nu}^3$, which have been calculated assuming the BPL (left panel) and H3a (right panel) models for the primary cosmic ray flux. We have that the different models for the inelastic proton - Air cross section predict distint values for the neutrino flux in the kinematical range currently covered  by the IceCube observatory ($E_{\nu} \lesssim 3 \times 10^6$ GeV),  with the MFEK prediction being a lower bound and the QGSJETII.03 the upper one. Such result is expected from the analysis of the Eq. (\ref{eq:zpH}), which show that the $Z_{pH}$ moment is inversely proportional to $\sigma_{inel}^{pAir}$. A surprising aspect is the fact that the EMM prediction is the second smaller one. From the analysis of the Fig. \ref{fig:1} at high energies, we will expect that the resulting flux would be the larger one and similar to the QGSJETII.03 one. However, we should to take into account that in the cascade evolution, the contribution of small energies is also important. As the EMM prediction overestimate the data at low energies, it reduces the value of the moments and implies a smaller prediction of the neutrino flux. Such result indicates that the modelling of $\sigma_{inel}^{pAir}$ should also be under control at low energies in order to obtain realistic predictions for the neutrino flux. At very large neutrino energies (beyond the IceCube energies), the predictions of all models, excluding the MFEK one, are similar for both primary fluxes considered. In order to estimate the magnitude of the difference between the predictions, in Fig. \ref{fig:ratio} we present our results for the ratio between the prompt neutrino flux predicted by the distinct models and that obtained considering the   
QGSJETII.03 one. The predictions are almost independent of the primary cosmic ray flux considered. We have that the two different versions of the QGSJET model differ by approximately 50 \% at low energies. On the other hand, the EMM one predicts a flux that is smaller than the QGSJETII.03 by $\approx 40\%$. Finally, the QGSJET01 and MFEK predictions differ by $\approx 3$ in the kinematical range covered by the IceCube Observatory. Such results indicate that the modelling of the inelastic proton -- Air cross section is an important source of uncertainty in the predictions of the prompt neutrino flux. Certainly, a more precise measurement of $\sigma_{inel}^{pAir}$ will be useful to constrain the description of the hadronic interactions in the cascade evolution equations.

Finally, let's summarize our main results and conclusions. One of the main shortcomings on the  determination of the atmospheric neutrino flux, as many other observables related to cosmic ray physics, is associated to the fact the predictions for the observables are strongly dependent on the modelling of the strong interactions at high energies. In particular, the prompt neutrino flux at the IceCube Observatory and future neutrino telescopes  depends on our knowledge about the hadronic interactions at the relevant energies, as well as several quantities, such as the primary cosmic ray spectrum, the longitudinal momentum distribution ($x_{F}$) of the incident particles and the inelastic proton-Air cross section. In this paper we have complemented previous studies and investigated the dependence on the modelling of $\sigma_{inel}^{pAir}$. We have considered some examples of phenomenological models that are largely used in the literature and demonstrated that the corresponding predictions for the prompt neutrino flux can differ by a factor of 3 in the kinematical range covered by the IceCube Observatory. Such result demonstrate that a future precise measurement of the inelastic proton - Air cross section is fundamental in order to derive realistic predictions of the prompt neutrino flux.

\begin{acknowledgments}
This work was  partially financed by the Brazilian funding agencies CNPq,  FAPERGS and  INCT-FNA (process number 464898/2014-5).
A.V.G. gratefully acknowledges the Brazilian funding agency FAPESP for financial support through grant 17/14974-8.
\end{acknowledgments}

\hspace{1.0cm}

\end{document}